\documentclass[aps,showpacs,showkeys,superscriptaddress]{revtex4-1}

\usepackage{graphicx}
\usepackage{epsfig}
\def\be{\begin{equation}}
\def\ee{\end{equation}}
\def\ba{\begin{eqnarray}}
\def\ea{\end{eqnarray}}
\def\half{{1 \over 2}}

\def\g5{\gamma_5}
\def\oper{{\cal O}}

\begin{document}

\def\wu{\widetilde{u}}
\def\wv{\widetilde{v}}

\title{ D-Brane solutions under market panic}

\author{R. Pincak}\email{pincak@saske.sk}

\affiliation{Institute of Experimental Physics, Slovak Academy of Sciences,
Watsonova 47,043 53 Kosice, Slovak Republic}
\affiliation{Bogoliubov Laboratory of Theoretical Physics, Joint
Institute for Nuclear Research, 141980 Dubna, Moscow region, Russia}
\date{\today}

\begin{abstract}
The relativistic quantum mechanic approach is used to develop a
stock market dynamics. The relativistic is
conceptional here as the meaning of big external volatility or volatility shock on a financial
market. We used a differential geometry approach with the parallel
transport of the prices to obtain a direct shift of the stock price
movement. The prices are represented here as electrons with different
spin orientation. Up and down orientations of the spin particle are
likened here as an increase or a
decrease of stock prices. The paralel transport of stock prices is
enriched about Riemann curvature which describes some arbitrage opportunities in the
market. To solve the stock-price dynamics, we used the Dirac equation for
bispinors on the spherical brane-world.
We found that when a spherical brane is abbreviated to the disk on the equator, we converge to the ideal behaviour of financial market where Black Scholes as well as semi-classical equations are sufficient.
Full spherical brane-world scenarios can descibe a non-equilibrium market behaviour were
all arbitrage opportunities as well as transaction costs are take into account.

\end{abstract}


\maketitle

\section{Introduction}

It has often been argued, and there is corroborating empirical evidence which suggests that stock market prices exhibit wave like properties~\cite{A0,B0}.
In the classical financial mathematics there were performed fundamental investigations to find  adequate stochastic processes matching the real financial data: Brownian, geometric Brownian, general Levy processes. From the point of view of quantum-like approach the problem cannot even be formulated in such a way. There is no any classical stochastic process which will match with the real financial data, because there is no a single Kolmogorov space describing the whole financial market. The financial data can only be represented as a quantum-like financial process~\cite{C0}.

The instantaneous return on the Financial Times-Stock Exchange (FTSE) and all Share Index is viewed as a frictionless particle moving in a one-dimensional square well (potential barrier) but where there is a non-trivial probability of the particle tunneling into the well’s retaining walls with some prices uncertainty~\cite{A,B,C,D}. The potential barrier approach is frequently used for barrier option calculations in the stock market~\cite{Jana}.
The wave like characteristic of the stock market prices gives rise to also quantum Nash equilibrium to solvable games in the Hilbert space~\cite{D0}.

Combinatorial auctions on financial game theory seem to be the most promising field. The promising quantum like experiments give rise to commercial implementation of quantum auctions in near future~\cite{F}.
The quantum model which gives surprisingly quantitative and transparent explanation of the observed shape and temporal stability of the observed stock price fluctuations on short time scales was developed in~\cite{FF}.
By including sufficiently many hidden variables stochastic models may be able to reproduce the historic observations with similar accuracy. Indeed, the (possible) existence of hidden variables was at the heart of the early critique of quantum theory. It is almost self-evident that the return on a stock depends on many factors that have not been modelled. The question nevertheless is not just one of having a more efficient description of the dynamics. Unlike hidden variables describing physical phenomena, the factors that influence the dynamics of a stock are expected to change over time. It furthermore is not clear how economic factors like the gross national product influence the value of any given stock at any given point in time. The observed scaling of the return distributions for various stocks in different economic environments strongly suggests that all these ”hidden” factors find their expression in the mean return and the variance of the returns.
The dynamics that determines the shape of the return distributions, on the other hand, must be self-consistent and largely immune to the influence of ”hidden” variables that are specific to a company and the economic and political climate. Hidden variables in our model  is an analogy with hidden dimensions in string and brane theory, both are non-visible non-measurable but have a big influence on the whole system.

The fluctuations in the real market are not Gaussian \cite{liz} unlike it does not decay as fast as Gaussian, and the history shows that the catastrophic loss (gain) is more likely than the Gaussian model; therefore some attempt to define non-linear quantum like model were made in~\cite{Yu,Ivanc,rac}.
It was found in~\cite{carf, ilin} that the space of
financial events is a fibration endowed with a particular affine
connection, so they are consequences of purely geometric properties.
The curvature of manifolds is determined by the connection upon the
fibration. The state space is considered as a smooth manifold the use of
rieman metrics and connections. In this assumption, we rewrite the known
pricing equations in terms of geometrical covariant quantities. We
defined  another type of geometrical object on financial market, which
may live on Riemannian manifolds, the spinor field. 

Some quantum mechanical approach to the option pricing dynamics was
firstly popular by described in~\cite{quant,quant1}. The option prices
in our model are represented by the electrons with mass
$M=1/\sigma^{2}$, where sigma is volatility, as in work~\cite{cont}.
In contrast to other quantum models with ordinary
particles~\cite{Haven,Haven1,Haven2,Choust,Choust1,Melnyk}, we assume an
electron with spin up and down which entitles us to use some
relativistic approach. Moreover, difference in spin orientation for
one electron could play a key role in difference for up and down factors of
increasing and decreasing option price in one step time interval of
change price in discrete binomial and trinomial models~\cite{Hoek,Tian}. In mathematical
definitions, the prices are represented via spinors as
elements of spaces in which these representations act as matrices, so
spinors will be ordinary real column vectors of prices in our model.
With the concept of spin bundle the spin connection will be calculated where
in terms of this connection spinors are parallel transported and
covariantly differentiated. 

It is clear that call and put prices are
moving on hyperbolic curves~\cite{hyperbol,hyperbol1} with a slope very similar to spherical curves in some cases.
Perhaps the simplest type of compactification leading to the existence
of chiral fermions is related to orbifolds. Consider a 3d
space-time of the form $S_1\times O\times R_1$, where $R_1$ is
(non-compact) time, $S_1$ corresponds to large observable
dimension with size $L$ ($0<x\leq L$), and $O$ is a (short)
interval corresponding to extra dimension  ($-R/2 \leq z \leq
R/2$), $L\gg R$. We will describe the dynamics of the relativistic particle which decsribe in this model the price of the stock market 
by Dirac Equation~\cite{greiner}.
The Dirac equation $\hat{D}\psi=E\psi$ for the 3-dimensional
two-component fermions
\begin{equation}\hat{D}\psi=\label{1}i\gamma^{\alpha}e_{\alpha}^{\mu}\nabla_{\mu}\psi,\end{equation} 
with
\be
\Psi(t,x,z) = \left( \begin{tabular}{l} $\psi_1$\\ $\psi_2$
\end{tabular}  \right) \;
\ee
has the form
\be
i \gamma^A \partial_A \Psi + M(z)\Psi =0~.\label{dirac}
\ee
$M(z)$ is a mass term which in general depends on the extra
coordinate $z$. Now we will specialize in the market panic or market under stress.
In~\cite{panic}, it was proposed that the drastic change in the cross-sectional dynamics of markets in a panic situation can be modeled as a phase transition induced by an external volatility shock.
The external volatility shock $\sigma_{ext}^{shock}$ is global or market wide, as would be
the case from an exogenous news event, then the individual internal volatilities $\sigma_{int}$ of each stock will become more similar, which leads cross-sectionally to a more Gaussian distribution, and low kurtosis.
The market volatility $\sigma_{M}$ under market panic can increase to values larger than critical volatility $\sigma_{c}$ due to either i) exogenous jumps (news) affecting all stocks so that $\sigma_{int}$ becomes $\sigma_{int} +\sigma_{ext}^{shock}$, or ii) endogenous, idiosyncratic jumps which are more stock-specific and can, for example, be induced by a large negative return.
The market panic in our approach is defined 
$M(z)=1/(\sigma_{ext}^{shock})^{2}\approx 0$, because the external volatility shock is more bigger than internal, 
e.q ($\sigma_{ext}^{shock}=40\%$ and $\sigma_{int}=20\%$ as in~\cite{panic}).
The bigger volatility is the higher velocity of change the stock prices. With the bigger market velocity the stock price particle  comes to the relativistic behaviour with small mass.

Moreover, we can image that an extra brane coordinate $z=\sigma_{int}$ is some kind of internal (implied) or local volatility~\cite{leb}, or moreover augmented volatility with transaction costs taken into account, see~\cite{taleb}. In this case the system switches on into the critical dynamics where external volatility shock $\sigma_{ext}^{shock}$ is big, fluctuates and is very difficult to predict. That behaviour in financial dynamic means very nonstable
financial market panic. In such extremal and economical crisis behaviour of the financial market to described dynamic of market panic some another more sensitive approaches need to be used.

\section{Model of the spherical brane markets}

\begin{figure}
\epsfysize=5.5cm \centerline{\epsffile{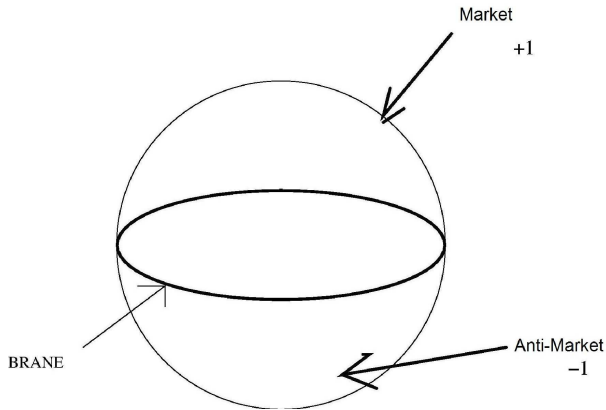}}
\caption{A manifold with the topology of a sphere and a domain
wall along the equator. +1 and -1 symbolically represent the different markets with particles and antiparticles on the opposite sites. The brane represents the domain of the ideal market where Black Scholes as well as semi-classical models work. } 
\label{fig:sphere}
\end{figure}

The basic mathematical construction for this paper are works~\cite{brane,Pincak1}, and we would like to utilize these ideas fo the financial market.
We suppose that the non equilibrium financial market is characterised by spherical brane-world
and equilibrium ideal market with zero arbitrage and transaction costs live on a domain 
wall along the equator, see Fig.\ref{fig:sphere}. 
The sphere model of financial economics encompasses finance, micro investment theory and much of the economics of uncertainty. As is evident from its influence on other branches of economics including public finance, industrial organization and monetary theory, the boundaries of the sphere are both permeable and flexible. 
Our considerations are based on recent publications.
Recently, in~\cite{sphere1} the brownian random walk and  diffusion
process on the sphere were developed. Also in~\cite{sphere2}, the quantum sphere model for a stock was introduced.
Moreover, work~\cite{sphere3}  shows that the risk-neutral world of the quantum binomial markets exhibits an intriguing structure as a disk in the unit ball of $R^{3}$, whose radius is a function of the risk-free interest rate with two thresholds which prevent arbitrage opportunities from this quantum market. 
In this work, the process of trading in a stock market was ilustrated by means of an elastic sphere model which is contextual and has a non-Kolmogorovian quantum like structure. As a consequence, for some theoretical support of the quantum models for the stock market.
We extend these ideas to the spherical brane-world of the financial market.

An alternative way to obtain chiral fermions or stock prices from extra dimensions
is to consider a (2+1)-dimensional theory for which the space is a
2d sphere. 
The sphere with radius $r$ of $S^2$ may be parameterized by two
spherical angles $q^1=\theta$, $q^2=\phi$ that are related to
Cartesian coordinates $x,\, y,\, z$ as follows:
\begin{equation} \label{sph-coord}
 x = r\sin \theta \cos \phi ;   \quad
 y = r\sin \theta \sin \phi ;   \quad
 z = r\cos \theta .
\end{equation}

The nonzero components of the metric tensor for sphere are
\begin{equation}
g_{\phi\phi}=r^{2}\sin^{2}\theta; \quad
g_{\theta\theta}=r^{2},\quad
\label{eq:5}
\end{equation}
where~$a,c\geq0,~0\leq\theta\leq\pi,~0\leq\phi<2\pi$.
Accordingly, orthonormal frame on spheroid is
\begin{equation}
e{_{1}}=\frac{1}{r~\sin\theta}~\partial_{\phi};\quad
e{_{2}}=\frac{1}{r}~\partial_{\theta}
\label{eq:6}~,
\end{equation}
and dual frame reads
\begin{equation}
e{^{1}}=r~\sin\theta~d\phi;\quad
e{^{2}}=r~d{\theta}\label{eq:7}
~.
\end{equation}
A general representation for zweibeins is found
to be
\begin{equation}
e{^{1}}_{\phi}=r~\sin\theta;\  e^{1}_{\ \theta}=0;\ e^{2}_{\
\phi}=0;\
e{^{2}}_{\theta}=r
\label{eq:8}
~.
\end{equation}
Notice that $e^{\mu}_{\ \alpha}$ is the inverse of
$e^{\alpha}_{\ \mu}$. The Riemannian connection with
respect to the orthonormal frame is written as~\cite{Nakahara,Gockeler}
\begin{equation}
de^{1}=-R^{1}_{\ 2}\wedge e^{2}=
\cos\theta\
d\phi\wedge e^{2},
\label{eq:9}
\end{equation}
\begin{equation}
de^{2}=-\omega^{2}_{\ 1}\wedge e^{1}=0.
\label{eq:10}
\end{equation}
Here $\wedge $ denotes the exterior product and $d$ is the
exterior derivative. From equations above we get the Riemannian connection
in the form
\begin{equation}
R{_{\phi 2}^{1}}=-R{_{\phi 1}^{2}}=\cos\theta; \quad
R{_{\theta 2}^{1}}=R{_{\theta 1}^{2}}=0.
\label{eq:11}
\end{equation}
The paralel transport of prices on the financial market was well described in \cite{spin}. The meaning of the paralel trasport was there described as exchange rates between different currency. 
The spin connection term in our apprach describes exactly some arbitrage opportunity in the stock market. 
In work \cite{arb}, an option pricing model with endogenous stochastic arbitrage was developed, that is capable of modelling in a general fashion any future and underlying asset that deviate itself from its market equilibrium. They found that the consequences of a finite and small endogenous arbitrage not only change the trajectory of the asset price during the period when it started, but also after the arbitrage bubble has already gone. The new trajectory of the stock price was analytically estimated for a specific case of arbitrage. The different trajectories from the initial to the final point in the language of differential geometry means the non zero curvature.
Therefore, arbitrage opportunities are closely connected with curvature. If the curvature is zero the stock market has also zero arbitrage opportunity, see \cite{spin2}, it means that there is only one trajectory, possibility how to do some trade or operation on the financial market. Moreover, in Ref.\cite{spin2} it was found empirically from the financial data that arbitrage has a tendency to oscilate on the financial stock market. This is also the result from our approach where Eg.(11) exactly proves this behaviour.  

Spontaneous symmetry breaking generated by the control parameter imposes a phase transition between the arbitrage phase and no-arbitrage phase developed in \cite{spin3}. In contrast to the short-living mode which is expected in the frame of the efficient market hypothesis, the long-living modes are totally new and exotic. The existence of the spontaneous arbitrage mode explains why the arbitrage return survives longer than expected and why the trading strategies based on market anomalies can make profits for a long time.

Spinors(prices) in two dimensions have two components and the role of Dirac matrices
belongs to Pauli matrices: $\gamma^a \rightarrow (\sigma_x,\, \sigma_y, \,\sigma_z)$
where
\begin{equation}\label{Pauli}
  \sigma_x = \left(
\begin{array}{cc}
  0 & 1 \\
  1 & 0
\end{array} \right); \qquad
\sigma_y = \left(
\begin{array}{rr}
  0 & -i \\
  i & 0
\end{array} \right); \qquad
\sigma_z = \left(
\begin{array}{rr}
  1 & 0 \\
  0 & -1
\end{array} \right).
\end{equation}

Covariant derivatives of 2-spinors are also expressed in terms of the spin
connection:
\begin{equation}\label{Dpsi}
  \nabla_\alpha \psi =
    \partial_\alpha \psi + \frac i4 R_\alpha^{ab}\, \sigma_{ab}\, \psi.
\end{equation}
Here $\sigma_{ab}$ are the rotation generators for spin-$\frac 12 $ fields:
\begin{equation} \label{sigma12}
  \sigma _{1\,2} = - \sigma _{2\,1} =
    - \frac i2 [\gamma_1,\, \gamma_2] = \sigma_z.
\end{equation}
A standard calculation gives for covariant derivatives:
\begin{equation} \label{nablaS2}
 \nabla_\theta  =  \partial_\theta
    \qquad \mathrm{and}  \qquad
  \nabla_\phi  = \partial_\phi - \frac {i\sigma_z}2 \cos \theta.
\end{equation}

Now we may define the Dirac operator. In spherical coordinates it is given
by the convolution of covariant derivatives in the spinor representation with
the zweibein and $\sigma$-matrices:
\begin{equation}\label{Dirac-S2}
  -i \hat{\nabla} = -i\, e^{\alpha\,a} \sigma_a \nabla_\alpha.
\end{equation}
With the help of the above definitions it
is straightforward to obtain:
\begin{equation}\label{Dirac/sph}
  -i\,\hat{\nabla} =
    -i \sigma_x
    \left(\partial_\theta + \frac{\cot\theta}2 \right)
    -\frac{i \sigma_y}{\sin\theta}\, \partial_\phi .
\end{equation}
Here $\theta$ and $\phi$ are the usual polar coordinates on the
sphere.
Consider the square of  the Dirac
operator $(-i \hat{\nabla})^2$. Obviously, if $-i \hat{\nabla}$ has
a zero eigenvalue, then $(-i \hat{\nabla})^2$ must have it either.
Let us split the product of $\sigma$-matrices in
$(\hat{\nabla})^2$ into symmetric and antisymmetric parts
 $\sigma^\alpha \sigma^\beta = \half \{\sigma^\alpha,\, \sigma^\beta\} +
    \half [\sigma^\alpha,\, \sigma^\beta]$.
Implementing the relation between commutators of covariant derivatives and
curvature we obtain:
\begin{equation}\label{D-L=comm}
    (-i \hat{\nabla})^2 + \nabla_F^2 =
    -\frac i2 \sigma^{\alpha \beta}\, [\nabla_\alpha , \, \nabla_\beta] =
    \frac 14 \, R^{\alpha\beta}_{\alpha\beta}.
\end{equation}
Here $R^{\alpha\beta}_{\alpha\beta}$ is the trace of the Riemann curvature
tensor and $\nabla_F^2$ is the covariant Laplace operator in the fundamental
representation of the $SU(2)$-group:
\begin{equation}\label{Laplace}
  \nabla_F^2 = g^{\alpha\beta}\,
    (\nabla_\alpha \nabla_\beta - \Gamma^\gamma_{\alpha\beta}\nabla_\gamma)
    = \frac 1{\sqrt{g}}\,
    \nabla_\alpha\, g^{\alpha\beta}\sqrt{g}\, \nabla_\beta ,
\end{equation}
where $\nabla_{\alpha,\, \beta,\, \gamma}$ are covariant derivatives and $\Gamma^\gamma_{\alpha\beta}$ is
the Christoffel symbol, and $g$ is the determinant of metric,
 $g = \det \|g_{\alpha\beta} \| = r^2\sin^2 \theta$.
In work  \cite{spin1} the same covariance of the Laplace operator on the curvilinear coordinates for stock price dynamics was described.
In the case
of the sphere with radius $r$ the rici tensor or curvature is
\begin{equation}
 R^{\alpha\beta}_{\alpha\beta} = 2/r^2,
\end{equation}
which is exactly the case of nonzero arbitrage possibility decribed above.

The problem has translational symmetry with respect to time and
the azimuthal angle $\phi$, so we can take the spinor to depend on
these as $\exp[-iEt+i m\phi]$, where $m=\pm \half, \pm \frac{3}{2}, \ldots$ is a
half-integer.  The parameter $m$ may be called the projection
of angular momentum onto the polar axis. In the financial brane market the parameter could describe some trends  where sign $\pm m$  indicates if future trends are with increase or decrease of stock prices
similar to a Parkinson number,  see~\cite{taleb}. It could be some inertia in rotation of fermion particle (stock price)
in a financial brane market.
We then obtain the following
equations for the components:
\ba
\left[ \partial_\theta + \half \cot \theta + {m\over \sin\theta}
\right] \psi_2
+ \Phi \psi_1 & = & \tilde{E} \psi_1 \; , \label{comp1} \\
\left[ \partial_\theta + \half \cot \theta - {m\over \sin\theta}
\right] \psi_1 + \Phi \psi_2 & = & - \tilde{E} \psi_2 \; ,\label{comp2}
\ea
where $\Phi$ is a scalar field, whose dependence on $\theta$
is for a moment arbitrary but for the next consideration it will be
a domain wall localized on the equator (i.e., at $\theta= \pi/2$).
From a financial point of view the scalar field $\Phi$ is some kind of arbitrage buble.
The stock market in our model is similar to a charged particle moving in the electromagnetic field, where the difference is that the external field of the stock market is constructed by the information. The stock price may be influenced by such a field. 

In~\cite{man}, the Yang-Mills function of financial markets obtained significant conserved quantity and derived financial solitons by transforming the correspondent space-time and non-Abel localization gauge symmetry. It shows that there is strict symmetry between the manifold fiber bundle and the gauge field on financial markets, that the differentiable manifold theory was preferably described the relation of internality interactions in finance markets (securities, futures). The characteristic of availability tracking markets price fluctuations explains that the conserved quantity is the traveling wave solution (soliton) of stock and futures price fluctuations. The financial solitons discovered indicate that there is a kind of new substance and form of energy existing in financial trade markets, which is the emergence of interactions under complex systems.
Therefore, $\tilde{E}$ is a, energy operator of the market where a one part can be devoted to the kinetic 
energy of the stock return, which represents some properties of the stock itself. 
From the kinetic part of energy we can defined the kinetic volatility component, as was shown in~\cite{gon}, where the interplay between the economic chaos and the volatility dynamics was explained. 
The volatility is also closely connected with the amplitude of the behaviour of the financial market, see~\cite{pincak}. Accordinly, from the experimental measurement of the changes of volatility from real financial data we could also find the changes of the kinetic energy of the market.
The next part of the energy operator corresponds to the potential energy reflects, e.q., the cyclical impact the stock feels in the information field or some interaction between market and investors, see~\cite{quant2}. We consider substitution $\tilde{E}=r E$ on the financial stock market where radius $r$ is function of risk-free interest rate~\cite{sphere3} which compensate in some sense an arbitrage opportunities. $E$ is the operator of energy in critical case of market panic and will be defined later. With bigger radius (interest rate) of spherical brane  the stock market energy is also bigger. Moreover, the curvature expression (20) reveals the following behaviour of the financial market:  with increasing interest rate (we have dear money with smaller market liquidity), the curvature decreases and also abritrage possibilities are slower, and we converge to the ideal market behaviour. The ideal market behaviour with non-arbitrage is in fact the case of principle of minimal variety in cybernetics~\cite{ash}. With incerasing volatility or amplitude of prices dispersion, the variety of market also increases and moreover if all subset of variety are oscilating with the same amplitude we come to the resonance behaviour of the market or market crisis where presented model could be suitable.
￼
￼￼

The Eqs.(21,22) form the eigenvalue problem for the Dirac operator
\be
\hat{D} = \left( \begin{tabular}{lr}
$\Phi$ & $\partial_\theta + \half \cot\theta + {m \over \sin\theta}$ \\
$-[\partial_\theta + \half \cot\theta - {m \over \sin\theta} ]$
& $-\Phi$
\end{tabular} \right) \; ,
\ee
whose square is
\be
\hat{D}^2 = \left( \begin{tabular}{lr} $\Phi^2 - [\partial_\theta^2
+ \cot\theta \partial_\theta - \frac{m^2 - m\cos\theta + {1\over
4}}{\sin^2\theta} - {1\over 4}]$ &
$-\partial_\theta\Phi$ \\
$-\partial_\theta\Phi$ & $\Phi^2 - [\partial_\theta^2 + \cot\theta
\partial_\theta
 - \frac{m^2 + m\cos\theta + {1\over 4}}{\sin^2\theta} - {1\over 4} ]$
\end{tabular} \right) \; .
\label{O2}
\ee
In~\cite{nonhermit},  it is also shown that, the non-Hermitian Hamiltonians of quantum theory may be related to the so-called quantum finance Hamiltonian.
We now see that the problem becomes particularly simple for $\Phi$
of the form of a step-function ($\Phi_0 > 0$):
\be
\Phi(\theta) = \left\{ \begin{tabular}{lr}
$\Phi_0 \; ,$ & $\theta < \pi/2 \; ,$ \\
$-\Phi_0 \; ,$ & $\theta > \pi/2 \; .$
\end{tabular}
\right. \label{Phi_step}
\ee
This corresponds to the limit of an infinitely thin domain wall.
This is a similar step arbitrage function as presented in work~\cite{cont}.
Moreover, we can go deeper and imaging that in our spherical brane 
in the hemisphere where $\theta <\pi/2$ we have defined the clasical financial market 
with the arbitrage scalar field $\Phi_0$. In the opposite hemispere when  $\theta >\pi/2$ we have defined 
some kind of antimarket with negative arbitrage or some kind of anti-arbitrage $-\Phi_0$. 
In upper hemisphere the market could be define by particle with possitive charge and currency pair EUR/USD in contrast to the lower hemisphere where the anti-market will be defined by particle with a negative charge and an opposite currency pair USD/EUR.
This is the behaviour of relativistic physics where from symmetry of Dirac equation there arise matter and anti-matter part
of the universe.
In the case of one step arbitrage, the eigenvalue equation for $\oper^2$ becomes
diagonal everywhere outside the equator, while at the equator the
off-diagonal terms in $\oper^2$ produce $\delta$-function
``potentials''. We adopt this choice of the scalar-field profile
in what follows. We can then use solutions for constant fermion
mass $\Phi_0$ and match them at the equator.
Solutions for constant mass can be expressed through
hypergeometric functions, using transformations described in \cite{Pincak1}. 
In what follows, we assume that $m > 0$.
Solutions for $m < 0$ can be obtained by reflection at the
equator. Define a new coordinate variable $z = \cos^2
\frac{\theta}{2}$, and a new pair of functions $\xi(z)$ and
$\eta(z)$:
\ba
\psi_1 & = & (1-x)^{\frac{m}{2} - \frac{1}{4}} (1+x)^{\frac{m}{2}
 + \frac{1}{4}} \xi \; , \\
\psi_2 & = & (1-x)^{\frac{m}{2} + \frac{1}{4}} (1+x)^{\frac{m}{2}
- \frac{1}{4}} \eta \; ,
\ea
where $x = \cos\theta = 2 z -1$. Then, the problem reduces to the
eigenvalue problem for the operator
\be
\left( \begin{tabular}{lr} $z(1-z) \frac{d^2}{dz^2} + [m + {3\over
2} - (2m+2) z] \frac{d}{dz} -a b$ &
$-(1-z) \Phi_{,z}$ \\
$-z \Phi_{,z}$ & $z(1-z) \frac{d^2}{dz^2} + [m + {1\over 2} -
(2m+2) z] \frac{d}{dz} -a b$
\end{tabular} \right) \; ,
\ee
where
\ba
a & = & m + \half + \sqrt{\tilde{E} ^2 - \Phi^2} \; , \\
b & = & m + \half - \sqrt{\tilde{E} ^2 - \Phi^2} \; .
\ea
By analogy with the financial market parameters we can denote $a=S$ as a stock price and $b=K$ 
as a strike price in the stock market. As it is clear from Eqs. (29,30), dependent on the 
energetics state of the market $\tilde{E}$, in comparison with arbitrage oportunities $\Phi$, it can be $S\leq K$ or $S \geq  K$.
In the case when $\tilde{E}=\Phi$  the $K=S$, strike price is equal to the stock price as in the case of options at the money.

For the scalar field (\ref{Phi_step}), we can construct the
eigenfunctions $(\xi, \eta)$ at $z \leq \half$ and $z \geq \half$ on both hemispheres 
from solutions to the hypergeometric equation that are regular at
the north pole  and the south pole, respectively. In our model extra coordinate $z=\sigma_{int}$. After solution of Eq. 24  we obtain functions of stock prices for the market and anti-market part of brane
\be
\xi = \left\{  \begin{tabular}{ll} $\mathcal{F}(S, K, m+\frac{3}{2}
; \sigma_{int})=\xi_1$ , 
& ~~~~~~~$\sigma_{int} \leq \half$ , \\
$\nu \mathcal{F}(S, K, m + \half; 1 -\sigma_{int})=\xi_2$ , & ~~~~~~~$\sigma_{int}\geq \half \; ,$
\end{tabular} \right. \label{xi}
\ee
and
\be
\eta =
\left\{  \begin{tabular}{ll} $-\alpha \nu \mathcal{F}(S, K, m+\half; \sigma_{int})=-\alpha \eta_1$ ,
& ~~~~~~~$\sigma_{int} \leq \half$ , \\
$-\alpha \mathcal{F}(S, K, m + \frac{3}{2}; 1 -\sigma_{int})=-\alpha \eta_2$ , &  ~~~~~~~$\sigma_{int} \geq \half \;
.$ \end{tabular} \right. \label{eta}
\ee
where $\mathcal{F}\equiv {}_2F_1$. 

The functions $\xi, \eta$ could be some analogies with call and put option prices respectively, or some derivatives of the stock prices with sensitivities in the internal volatility or in the spot FX rate as in Vanna-Volga methods~\cite{vanna}.
The similar solutions were found in~\cite{hyper} 
where after applying supersymmetric transformations on a known solvable diffusion process they obtain a hierarchy of new hypergeometric solutions for some class of superpotentials in Schrodinger equation. These solutions were given by a sum of hypergeometric functions, generalizing the results obtained in~\cite{hyperbol}  for two-dimensional processes.  
The parameter
\be
\nu = \frac{\mathcal{F}(S, K, m+\frac{3}{2}; \half)}{\mathcal{F}(S, K, m + \half;
\half)} \; .
\ee
From the jump of the derivatives on the equator, we obtain
$\alpha=\pm 1$ and the eigenvalue equation
\be
\frac{\nu \mathcal{F'}(S, K, m+\half; \half) + \mathcal{F'}(S, K, m+\frac{3}{2};
\half)}{4 \Phi_0 \mathcal{F}(S, K, m+\frac{3}{2}; \half)} =\alpha = \pm
1\; , \label{spec}
\ee
which determines the allowed energies $\tilde{E}$.
From the above equations we can find other special behaviour and characteristics of the option dynamics as Call-Put Duality and Call-Put Reversal,
as defined in~\cite{Jesper,Wilmott} . On the equator  where $z=\sigma_{int}=1/2$ we can directly write 
\begin{eqnarray*}
\xi_1=-\alpha \eta_2,\\
\eta_1=-\alpha\xi_2,
\end{eqnarray*}
where call and put options can be equal, dual or reversal.

In the limit applies in the case of main interest to
us: $\Phi_0 \gg 1$ and $\tilde{E} \ll \Phi_0$, which corresponds to a very big arbitrage buble, and variety of market
which is case of market panic.
This state has $\alpha =1$, and
for its energy in the case of market panic we obtain
\be
E^2 = m^2 / {r}^2 + O(m^2 /  {r}^2\Phi_0^2) \; , \label{E2}
\ee
which is the dispersion law of a
massless fermion in our case stock prices propagating along the equator. Dispersion in our approach means 
deviation from the ideal market defined on the equator domain.
The sign of $E$ can be found by returning to Eqs. (\ref{comp1}), 
(\ref{comp2}). We find $E\approx -m / {r}$, which corresponds to 
a left-moving, i.e., chiral fermions or prices in (1+1) dimensions in anti-market hemisphere.

For the time evolution of the above finding solutions we can write
\be
\Psi(\sigma_{int}, \phi; t) = \frac{1}{\sqrt{2\pi}} \sum_m e^{im\phi} \left(
\begin{tabular}{r} $\xi_m(\sigma_{int})$ \\ $\eta_m(\sigma_{int})$ \end{tabular} \right)
A_m(t) \;. \label{proj}
\ee
$A_m$ is the amplitude of a single-component (chiral) 2d
fermion in the interval $0\leq t \leq T$ where  $T$ is maturity time of call or put option.

\begin{figure}
 \begin{center}
 \includegraphics[height=8cm]{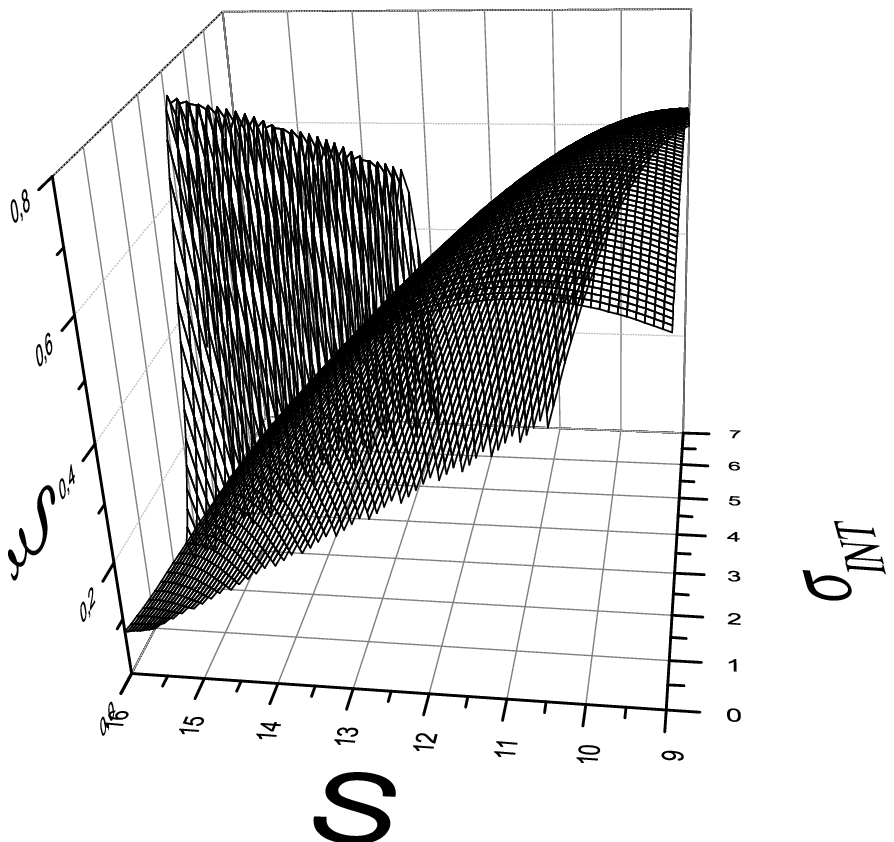}
 \caption{}
 \end{center}
 \end{figure}

\begin{figure}
 \begin{center}
 \includegraphics[height=8cm]{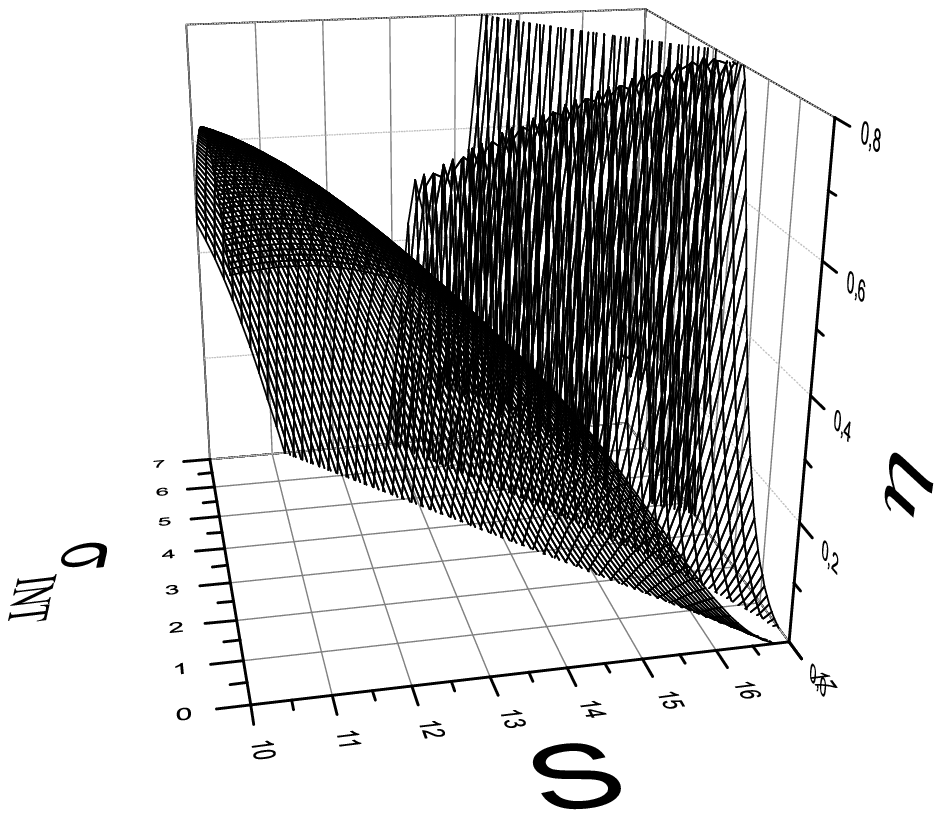}
 \caption{}
 \end{center}
 \end{figure}

\section{Stock market on a disk}

Our results of the stock market  on a sphere are related to the topology, rather than geometry,
of the spatial manifold. Similar results are valid for a simpler, flat
geometry. Simply cut a sphere along the equator, chose a
hemisphere, and make it flat (with zero curvature and therefore also arbitrage oportunities) by replacing it with a disk. Then,
substitute the domain wall by a suitable boundary condition. 
This is similar structure as a disk galaxie which is system in centrifugal equilibrium.
In this subsection we shortly present the corresponding equations. We will
call the boundary of the disk the brane and its interior the bulk.
Introduce the Cartesian coordinates $x$ and $y$ with the origin at the
center of the disk. Three-dimensional $\gamma$-matrices used in
this subsection are $\gamma^0=\tau_3,~\gamma^1=i \tau_1,~ \gamma^2
= i \tau_2$. Note that these matrices are associated with the
Cartesian coordinates. Then, the Dirac operator on a disk in the polar coordinates ($ x
= r \cos\phi,~y=r\sin\phi$)  is
\be
\hat{D}= \left( \begin{tabular}{lr}
$\Phi$&$e^{-i\phi}\left(-\partial_r+\frac{i}{r}\partial_\phi\right)$ \\
$e^{i\phi}\left(\partial_r+\frac{i}{r}\partial_\phi\right)$&$-\Phi$
\end{tabular} \right) \; ,
\ee
leading to the energy eigenvalue problem $\hat{D}\Psi =E \Psi$. The
regular solutions are:
\be
\left(\begin{tabular}{l}
$\psi_1$\\
$\psi_2$
\end{tabular} \right)=
\left(\begin{tabular}{l}
$~~~~~~~~~e^{in\phi}J_m(k r)E_+$\\
$-e^{i(n+1)\phi}J_{m+1}(k r) E_-$
\end{tabular} \right)
 \; ,
\ee
for $E^2>\Phi^2$ and
\be
\left(\begin{tabular}{l}
$\psi_1$\\
$\psi_2$
\end{tabular} \right)=
\left(\begin{tabular}{l}
$~~~~~~~e^{in\phi}I_m(k r)E_+$\\
$e^{i(n+1)\phi}I_{m+1}(k r) E_-$
\end{tabular} \right)
 \; ,
\ee
for $E^2<\Phi^2$. Here $E_+=\sqrt{|E+\Phi|},~~E_-=\sqrt{|E-\Phi|},~~k
=\sqrt{|E^2-\Phi^2|}$, $J_m$ and $I_m$ are the Bessel and modified
Bessel functions. The parameters $E, \Phi, r$ are energy operator, 
arbitrage buble, interest rate (radius of the sphere) and $m$ (projection of angular momentum) could be an indicator of some trends of the stock market. Similar results in the form of Bessel functions for the stock market were
found in~\cite{extbessel}. For the operator energy in the case of disk stock market one can find $E=  m / {r}$.

\section{Conclusions}
As was published many times, the Black Scholes model is good idealization of the financial market. 
It is known by referring to the experimental data published in many journals that the random walk model is a good approximation of the market reality in static situations or in a equilibrium state on the financial market. Whenever such extra-ordinary events take place especially leading to financial and/or economical crisis, the random walk model fails. It means in the case of very big  instability with very big volatility and fluctuation of prices that some other approaches need to be developed to describe and obtian real dynamics of such market panic behaviour. 

Market equilibrium comes at the price of a commodity for balancing the market forces like demand \& supply. In market equilibrium the amount that the buyers want to buy is equal to the amount that the sellers want to sell. The reason we call this equilibrium, when the forces of demand \& supply are in balance, there is no reason for a price to rise or fall as long as other factors remain unchanged. A situation in which the supply of an item is exactly equal to its demand. Since there is neither surplus nor shortage in the market, price tends to remain stable in this situation. Equilibrium price is also called market clearing price because at this price the exact quantity that producers take to market will be bought by consumers, and there will be nothing ‘left over’. For markets to work, an effective flow of information between buyer and seller is essential. The flow of information is described in our approach as a interaction part of the energy operator of market.

The equilibrium market behaviour could be obtained with two possibilities in our model. The  first one is that $E=\Phi$,  it means that arbitrage posibilities compensate the energy of the market. For example, interactions between investors and markets are so stable that there are no any arbitrage possibitites. The second possibility is the case of ideal market behaviour when arbitrage posibilities are zero from the beginning $\Phi=0$, and for the ideal behaviour of the market the energy of the market needs to be $E=0$. E.q. there are no any interactions between the market and traders, any changes in prices, any transaction costs are taken into account. This is exactly the case of the perfect market equilibrium postulated by the  Black Scholes model. 

In our approach of the stock market  on a spherical manifold which is close connected into the  topology of the spatial manifold itself the result is that there are not any possibilities for such ideal behaviour of market. It means that on the sphere does not exist any zero modes solutions any case of ideal market behaviour where $E=0$ as can be seen from Eq.(35) and moreover this is a general property of manifolds with positive
curvature known as the Lichnerowicz theorem \cite{Berline/Getzler/Vergne}.  

On the other hand such zero mode solutions arise directly from Eqs.(38,39) for the first case of equilibrium when $E=\Phi$ in the case of the disk stock market. For the special choice of boundary conditions which could represent some market behavior we can always found zero mode solutions on disk stock market. The result is that disk market is some idealization of the real nonlinear market behaviour on the spherical brane and converge to the Black Scholes solutions. Figures 2,3 shows the behaviour of the stock market or some derivatives for a special choice of the model parameters to obtain a results of stock prices similar as from the Black Scholes and semi-classical equations. 

Finally to obtain real financial dynamics of stock market full mathematical equations Eqs.(26,27) and Eq.(36) obtained on the spherical brane with appropriate boundary conditions need to be exactly solved. To do all these analyses is the main qoal for the future work.

George Soros in his book \cite{soros} underlined a fundamental difference between natural and social sciences.
The events studied by social sciences have thinking participants and natural phenomena do not. The participants’ thinking creates problems that have no counterpart in natural science. There is a close analogy with quantum physics, where the effects of scientific observations give rise to Heisenberg uncertainty relations and Bohr’s complementarity principle. Moreover, the behaviour of the stock market described with the wave function or in our case with the spinor wave function of the stock market carries more real information of a market. 

In the book George Soros philosophically compare the predictable behaviour (prediction of prices) in our approach regular modes of the wave function with unpredictable behaviour (psychological factors of people in the market) as irregular modes of the wave function of the financial market. All clasical and semiclasical teories described only regular modes or predictable behaviour of the stock market.  Also Soros statement about cognitive phenomena at the financial market included deeper knowledge of the financial dynamics.

This paper is some attempt where a new approach was used to obtain descriptions of the dynamics of market panic on financial world.  This is only first review of some new relativistic approach to stock market dynamics where for concrete realizations and applications on financial market a lot of analyses and computations on real financial data need to be done. It is also some challenge for the reader for those our approach is interesting to bring  ideas from the theoretical field to the real market applications. These ideas can be extended to higher dimensional branes of financial market.

\vskip 0.4cm ACKNOWLEDGEMENTS --- The work was supported by the Slovak Academy of
Sciences in the framework  at VEGA Grant No. 2/0037/13.

\end{document}